# Mechanical and adsorption properties of greenhouse gases filled carbon nanotubes

Daniela A. Damasceno[1], Henrique M. Cezar[2], Teresa D. Lanna[3], Alexsandro Kirch[4], Caetano R. Miranda[5*]

[1,3,4,5] Universidade de São Paulo, Instituto de Fisica, Rua do Matão 1371, São Paulo, SP 05508-090, Brazil

[2] Hylleraas Centre for Quantum Molecular Sciences and Department of Chemistry, University of Oslo, PO Box 1033 Blindern, 0315 Oslo, Norway

Corresponding author: *Caetano R. Miranda, email: crmiranda@usp.br

## Abstract

We investigate the mechanical and adsorption properties of single-walled carbon nanotubes (SWCNTs) filled with greenhouse gases through Grand Canonical Monte Carlo (GCMC) and Molecular Dynamics (MD) simulations using a recently developed parameterization for the cross-terms of the Lenard-Jones (LJ) potential. Carbon nanotubes interact strongly with $CO_2$ compared to $CH_4$, resulting in a $CO_2$-rich composition inside the nanotubes, with the proportion of $CO_2$ decreasing as the diameter of the nanotubes increases. Contrarily, the smallest nanotubes showed a more even balance between $CO_2$ and $CH_4$ due to gas solidification. The gas does not affect the mechanical response of the nanotubes under tension, but under compression, it presents a complex relationship with the loading direction, nanotube's diameters, chirality, and to a minor extent, the gas composition. Filled zigzag nanotubes showed to be more stable in the presence of fillers, giving the best mechanical performance compared to the filled armchairs. The study confirms carbon nanotubes as effective means of separating $CO_2$ from $CH_4$, presenting good mechanical stability.

# 1 Introduction

Single-walled carbon nanotubes (SWCNTs) have remained a topic of significant interest in various fundamental and applied science fields [1] since their discovery and synthesis [2–4]. Their unique properties [5,6] and remarkable capacity to host a variety of molecules [7–10] make them well-suited for energy storage [11], fluid transport [6], and gas capture and separation applications [12].

Regarding the adsorption in SWCNTs, a key point is that the molecules are highly confined in such systems, with most of the phenomena occurring at the nanoscale level. At this scale, several issues related to interfacial interactions may play a crucial role in the system's performance and fundamental properties [13,14]. Previous studies on filled nanotubes [15–17] and nanocomposites [18] have shown that the mechanical properties of a confined system depend on the number of guest molecules and how well these molecules are arranged within the host system. This dependence can either enhance or weaken the bucking and tensile strength properties, compromising the systems' mechanical performance.

Ajayan and Iijima [19] conducted the first filled carbon nanotubes (CNTs) experiments, followed by other studies to investigate the effects of guest molecules on the overall performance of CNTs [20–23]. Most mechanical experiments focused on bending and Young's modulus [7]. Computational modeling studies have better described filled CNTs. For instance, previous studies used molecular dynamics (MD) simulations to investigate the buckling behavior of filled SWCNTs with $C_{60}$, $CH_4$, Ne, nickel, copper, platinum, and bimetallic alloys [17,24,25]. Other studies explored the effects of copper [26], hydrogen [27], $C_{60}$ [28], and iron [29] fillers on the mechanical performance of SWCNTs under different mechanical loadings.

Despite the considerable advancements made in studying the mechanical properties of the filled CNTs through experiments and computational modeling, little is known about their mechanical properties filled with greenhouse gases, which are systems of great technological interest nowadays. Most of the studies focused only on adsorption properties [20,30–33]; thus, a thorough understanding of these systems' mechanical responses is crucial for gas capture and separation applications.

This work focuses on the confinement of $CH_4$, $CO_2$, and their mixture within SWCNTs. Initially, we employ Grand Canonical Monte Carlo (GCMC) simulations with our recently developed parameterization for the cross-terms of the Lenard-Jones

(LJ) potential [34], considering different nanotube diameters, ranging from 8.11 Å to 18.8 Å, and different chirality. The SWCNTs were coupled to reservoirs with compositions ($CO_2$%:$CH_4$%) 100:0, 80:20, 50:50, 20:80, and 0:100, and pressures of 100, 200, and 300 atm at 313.15 K. Then MD simulations were carried out to investigate the mechanical properties of filled SWCNTs under tensile and compressive strain. Based on the authors' knowledge, the effect of $CH_4$ and $CO_2$ filling on the mechanical response of armchair (*ar*) and zigzag (*zz*) SWCNTs with different diameters has not been addressed in the literature.

## 2 Computational methods
### 2.1 Interatomic potentials

To describe the atomic interactions, we employed the adaptive intermolecular reactive bond order (AIREBO) [35] for the nanotubes, the OPLS-AA [36] for the $CH_4$, and the EPM2 [37] for the $CO_2$. AIREBO is a widely used potential in MD simulations, which enables the description of intra and intermolecular interactions by coupling the reactive empirical bond order potential [38], Lennard-Jones (LJ), and the torsional interactions. Several studies have employed this potential to simulate carbon-based structures and predict their tensile strength properties [39–43], considering a modified cut-off function to avoid nonphysical failure behavior [44–46]. The selection of force fields for the gas molecules was based on a previous study on the transport properties of $CO_2$ and $CH_4$ [47]. In the simulations, the gas molecules were considered rigid bodies. To describe the intermolecular interactions, the LJ potential with newly optimized parameters for the $CO_2$-nanotube and $CH_4$-nanotube interfaces was considered [34]. The LJ parameters are presented in Supporting Information **Table S1.** The Lorentz-Berthelot rules were used to obtain the LJ parameters for the interaction between different molecule species.

### 2.2 Grand Canonical Monte Carlo

We consider SWCNTs of different diameters and chirality to investigate the effects of these attributes on adsorption. Namely, we studied the adsorption of gases with compositions ($CO_2$%:$CH_4$%) 100:0, 80:20, 50:50, 20:80, and 0:100 in (6,6), (7,7), (8,8), (10,10), (12,12), and (14,14) armchair nanotubes, and (0,10), (0,11), (0,14), (0,17), (0,21), and (0,24) zig-zag nanotubes. The SWCNTs were about 160 Å long. The nanotubes were placed in boxes with sides of 50 Å to avoid interactions between replicas due to periodic boundary conditions. All the individual models were previously minimized using their respective force fields before the simulations.

We performed simulations using the Widom insertion method to find the chemical potential corresponding to our target pressures of 100, 200, and 300 atm at 313.15 K [48]. For each reservoir composition among ($CO_2$%:$CH_4$%) 100:0, 80:20, 50:50, 20:80, and 0:100, we carried out NPT simulations for $6\times10^6$ MC steps considering 500 molecules in a periodic cubic box, considering the first $10^6$ MC steps for equilibration. Probabilities for the translation, rotation, and volume change moves were 48.5%, 48.5%, and 3%, respectively. At each 2000 step, 7000 Widom insertions were performed to calculate the chemical potential. The final chemical potential values were employed in GCMC simulations of the gas phase to confirm that the $CO_2$:$CH_4$ ratios were maintained. These chemical potentials were then used for the simulations of the SWCNTs.

For each thermodynamic condition, reservoir composition, and SWCNT model, we performed GCMC simulations to insert the gases in the SWCNTs. During these simulations, the SWCNTs were kept rigid, and restricted insertions within a cylinder with the radius of the nanotube were performed. This radius was estimated by calculating the distance between diametrically opposed atoms. Translation, rotation, insertion, and deletion moves were attempted with a probability of 25% each. Configurational-bias Monte Carlo moves were used for insertions, considering 16 trial positions for each insertion move. These simulations were carried out for up to $5 \times 10^7$ MC steps until the convergence of energy and number of molecules of each species. After the convergence, the average number of molecules of each species was used to perform NVT simulations for $10^7$ MC steps, with the final configuration of these simulations being the ones used as the input for the MD simulations detailed below.

All simulations considered the SWCNTs and gas molecules as rigid bodies and employed a cutoff radius of 12 Å, with the Ewald summation method for long-range electrostatics. These simulations were performed using Cassandra [49]. We also used a Monte Carlo method to estimate the accessible volume of the nanotubes by employing a method similar to the one used by [50], in such a way that Ne atoms were randomly inserted in the SWCNTs to probe their energy. If the energy is negative, the probe is considered inside the accessible volume and outside if otherwise. The probe atom is then removed and repeated $10^6$ times to obtain the accessible volume estimate.

### 2.3 Molecular Dynamics

MD simulations were performed with the LAMMPS (Large-scale Atomic/Molecular Massively Parallel Simulator) [51] package to evaluate the mechanical response of gas-filled SWCNTs under tensile and compressive strain. Periodic boundary condition in the tube axis direction (z-direction) is considered. Infinitesimal structural deformations were applied along the nanotube's length (z-direction) with a strain rate of 0.001 ps$^{-1}$ and 0.5 fs timestep.

Before the stretching process, we performed an energy equilibration during 20 ps followed by a pressure equilibration using the isothermal-isobaric (NPT) ensemble at 313.15 K and zero pressure during 30 ps. This thermalization time was enough for the gases and nanotubes to reach a relaxed state. The $CH_4$ and $CO_2$ molecules were considered rigid, while the nanotubes were flexible. The uniaxial compressive/tensile strain was applied through a canonical ensemble (NVT/sllod) at a temperature of 313.15K. Knowing that the thermostatting strategy in confined systems may affect some properties [52–55], we tested different options (see section S3 in the Supporting Information) and observed no major differences. For the reported simulations, we thermostat only to the nanotube walls.

## 3      Results and discussions

### 3.1 Molecular compositions inside SWCNTs as a function of pressure and reservoir

Using the GCMC simulations, we were able to observe the $CO_2$ and $CH_4$ molecules competing for the adsorption inside the SWCNTs. The ratio $CO_2$%:$CH_4$% in the reservoir differs from the ratio inside the SWCNTs. The stronger interaction between the $CO_2$ and the nanotubes, compared to the interaction of $CH_4$ and the nanotubes [34], makes the composition of the gases inside the SWCNTs to be, overall, $CO_2$ rich. Up to a 100% selection of $CO_2$ is achieved but, overall, at least 70% of the molecules adsorbed inside the SWCNTs were $CO_2$ molecules even for the reservoir containing 20% of $CO_2$. This is observed regardless of the reservoir composition, as shown in **Fig. 1**.

The exceptions are the smallest nanotubes, (6,6) and (0,10), for which the balance of adsorbed $CO_2$ and $CH_4$ is more uniform, and can even lead to a $CH_4$ selection, as observed in the 80% $CO_2$ 20% $CH_4$ reservoir simulations. The strong confinement in these two sizes leads to a melting point depression, as predicted by the

Gibbs-Thomson equation. This is seen by the analysis of the configurations, where we observe that the fluid inside the SWCNT solidifies. In this case, the interaction between gas molecules plays a more important role than in the case of larger nanotubes.

A slight increase in the nanotube diameter completely changes this picture, and for the (7,7) SWCNTs, almost 100% of the adsorbed molecules are of $CO_2$, regardless of the reservoir composition. This is also observed, to a smaller extent, for the (0,11) nanotube, with the $CO_2$ amount reaching the maximum values among the zig-zag nanotubes for this size. Starting from the (7,7) and (0,11) SWCNTs, the amount of $CO_2$ inside the nanotube decreases when the diameter increases. This behavior is expected since for a nanotube with infinite diameter, one would expect to obtain the same amount of $CO_2$ as in the reservoir.

We also compared how the reservoir pressure affects the adsorption in the same nanotubes. The results, also shown in **Fig. 1**, indicate the higher the pressure, the lower the amount of $CO_2$ in the adsorbed gas is. Even for the reservoir with the lowest amount of $CO_2$ (20%) case, the amount of $CO_2$ inside the SWCNT can be up to 98% at 300 atm.

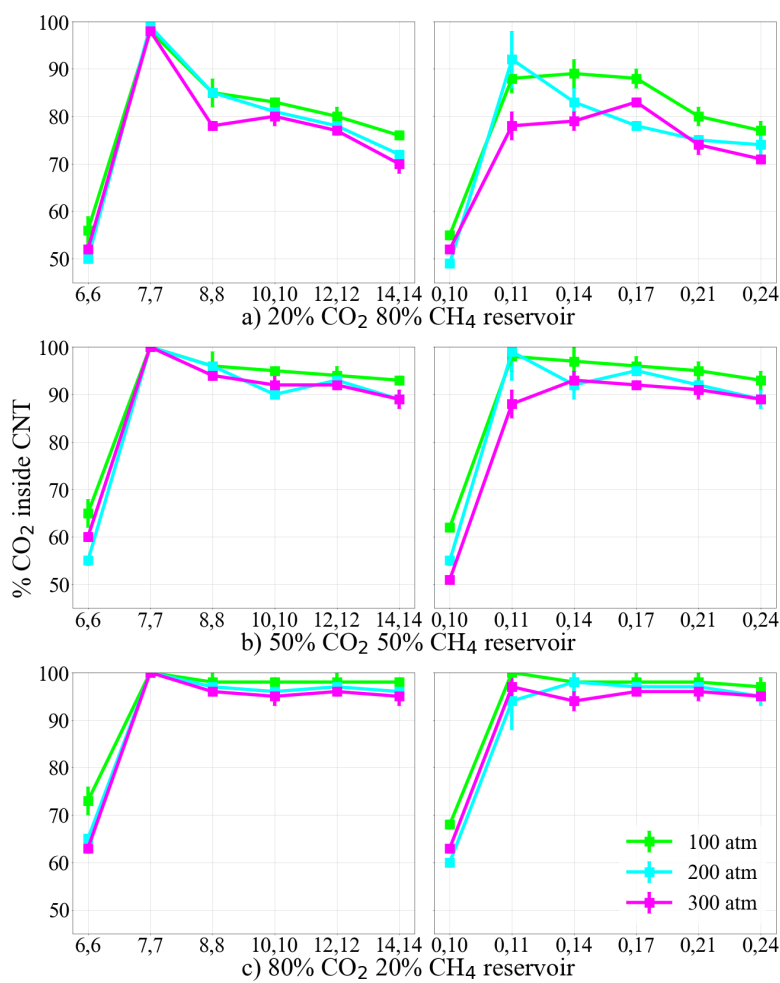

**Figure 1:** Percentage of $CO_2$ molecules inside the different CNTs at 100 (green), 200 (cyan), and 300 atm (magenta). Three different reservoir compositions were considered: (a) 20%$CO_2$:80%$CH_4$, (b) 50%$CO_2$:50%$CH_4$, and (c) 80%$CO_2$:20%$CH_4$. Observed compositions inside the CNT are different from those mixtures set in the reservoir due to the greater CNTs - $CO_2$ affinity.

We obtained the densities inside the SWCNTs considering the final number of adsorbed molecules for each reservoir, pressure, and nanotube diameter. The CNT accessible volumes were estimated using the Monte Carlo scheme as described in the methodology section. We observed the higher the confinement the more dense the gases inside the nanotube are (see **Fig. 2**). The densities display slight differences for the different pressures and the reservoir composition has a minor influence.

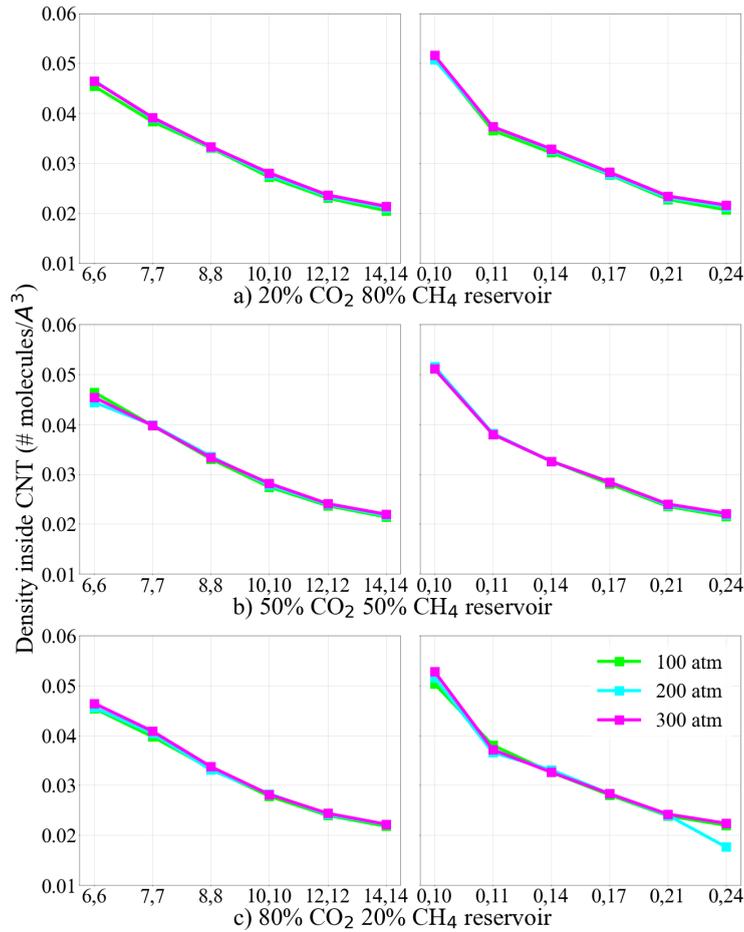

**Figure 2:** Densities of molecules inside CNTs for 100 (green), 200 (cyan), and 300 atm (magenta). Three different reservoir compositions were considered: (a) 20%$CO_2$:80%$CH_4$, (b) 50%$CO_2$:50%$CH_4$, and (c) 80%$CO_2$:20%$CH_4$. In the density estimation, we considered the total number of molecules ($CO_2$ and $CH_4$) per accessible volume.

### 3.2 Mechanical properties of SWCNTs filled with greenhouse gases

In **Fig. 3**, the fracture strength and failure strain of unfilled and filled *ar* and *zz* SWCNTs are presented along with their respective error bars. According to **Figure 3,** the fracture strength and failure strain of empty *ar* nanotubes range from 100.7 – 102.9 GPa and 0.197 – 0.202, respectively, while for *zz*, they range from 86.85 – 88.15 GPa and 0.131– 0.136. The values for the (12,12) and (0,21) nanotubes are consistent with those reported in the literature for nanotubes with a similar aspect ratio (~10) [56,57]. For the filled case, although a slight increase in the fracture strength and strain for the (8,8) is observed, most of the values fall within the empty case standard deviation. See **Fig. S2** for other compositions. In general, the gas molecules do not affect the mechanical response of the nanotubes under tension, regardless of the gas composition, chirality, and nanotube diameter. This finding is consistent with previous studies that observed similar behavior for empty and filled SWCNTs with n-butane and $C_{60}$ [7].

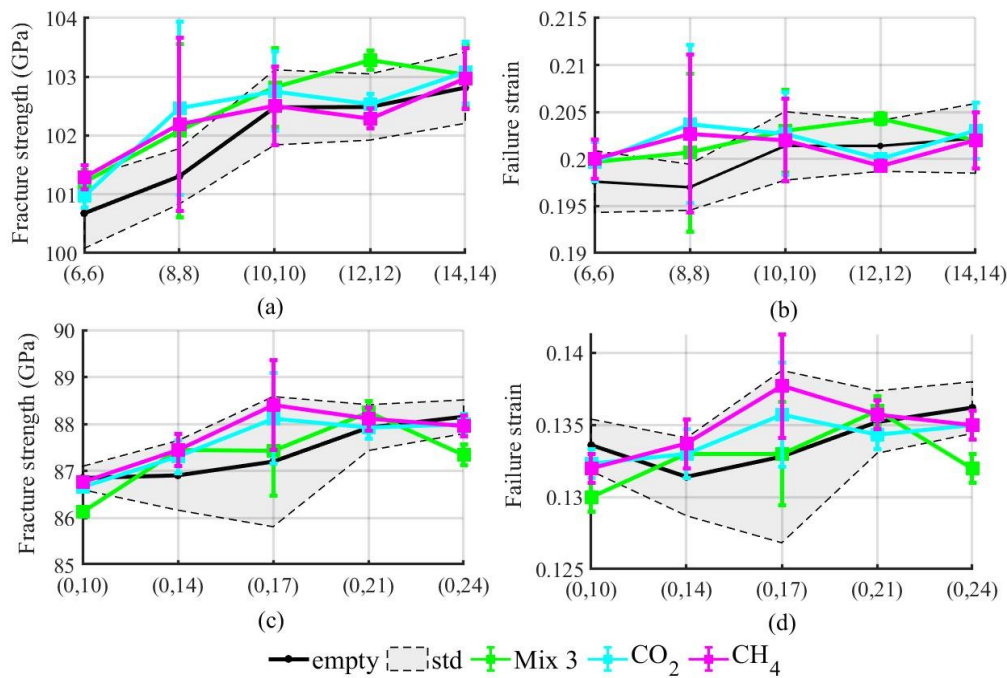

**Figure 3**: Comparison of (a) fracture strength and (b) failure strain of the empty and filled *ar* ((6,6), (8,8), (10,10), (12,12), and (14,14)); and (c) fracture strength and (d) failure strain of the empty and filled *zz* ((0,11), (0,14), (0,17), (0,21), and (0,24)) nanotubes. The mechanical response of empty nanotubes is represented by black curves and the gray region shows their standard deviation obtained based on an average of five different initial conditions. The green curves are for the composition Mix 3, shown in **Table S2** (See Supporting Information section S2), while the cyan and magenta curves represent the pure compositions of $CO_2$ and $CH_4$, respectively.

**Figure 4** presents the critical stress and strain for the configurations studied in the previous example (additional mix compositions are in the Supplementary material, **Fig. S3)**. We observed the critical stress increases with increasing diameter for both filled and unfilled nanotubes, up to diameter 13.56 Å (10,10) and 13.31 Å (0,17) inflection points. This trend is not chirality-dependent, but loading-dependent and suggests the existence of SWCNTs with a critical diameter for a given length, leading to a higher critical stress value. This behavior is in concordance with the one reported by previous studies that observed a similar trend in empty nanotubes [58,59].

Chirality effects, which were not significant among long empty nanotubes under compression [60], showed significance in the presence of fillers, particularly for *ar* nanotubes. **Figure 4** shows that larger nanotubes became stiffer with fillers, while the smallest one filled with pure gas composition became slightly more ductile. In the case of filled *zz* SWCNTs, the critical stress and strain remained in the same range as those of empty ones. Overall, filled *zz* nanotubes demonstrated greater stability in the presence of fillers under compression if compared to the filled *ar*, especially the (0,17) model. This behavior is similar to the one reported by previous studies on empty nanotubes [61], showing that the chirality and fillers may contribute to this effect on the mechanical response of nanotubes under compression.

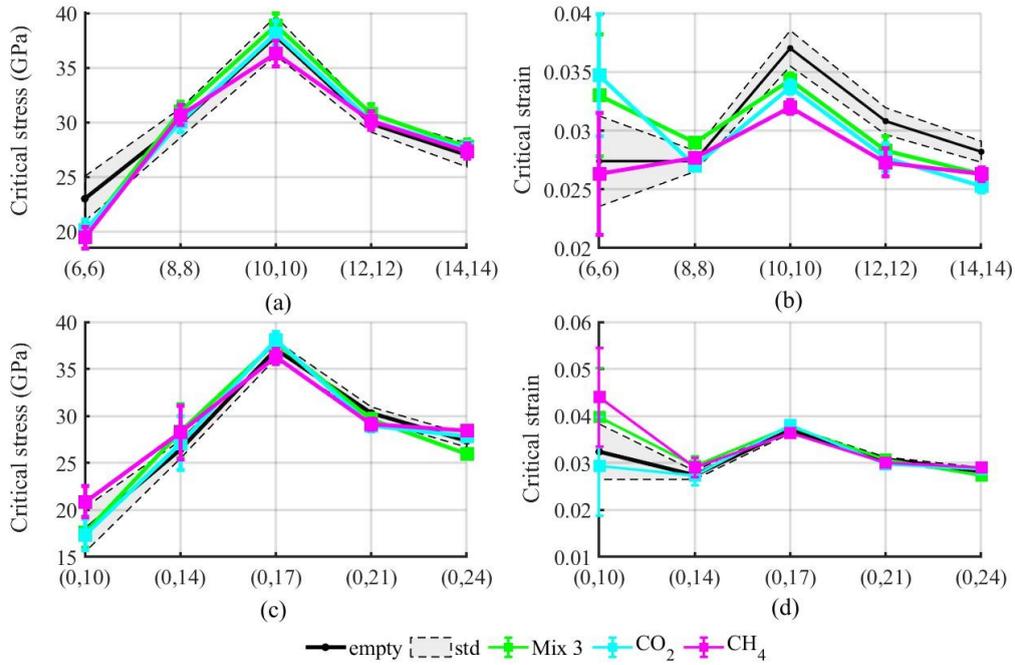

**Figure 4**: Comparison of critical (a) stress and (b) strain of the empty and filled *ar* ((6,6), (8,8), (10,10), (12,12), and (14,14)); and critical (c) stress and (d) strain of the empty and filled *zz* ((0,11), (0,14), (0,17), (0,21), and (0,24)) nanotubes. The mechanical response of empty nanotubes is represented by black curves and the gray region shows their standard deviation obtained based on an average of five different initial conditions. The green curves are for the composition Mix 3, shown in **Table S2** (See Supporting Information section S2), while the cyan and magenta curves represent the pure compositions of $CO_2$ and $CH_4$, respectively.

In addition, we investigated the filled nanotube mechanical response under three different reservoir pressures (100, 200, and 300 atm), as shown in **Fig. 5**. Our study examined the critical stress and strain of both *ar* and *zz* nanotubes under compression and tension strain for pure and mixed gas compositions (additional results are in **Fig. S4**). We observed the *ar* nanotubes were more impacted than the *zz* nanotubes under similar amounts of gases, resulting in a slight decrease in their critical stress and strain as pressure increased. However, the changes were not significant within this range of pressures. Nevertheless, previous studies on gas separation have reported a strong effect of pressure on other properties, such as permeability and selectivity [12,62], which should be considered when designing technologies for gas capture and separation membranes.

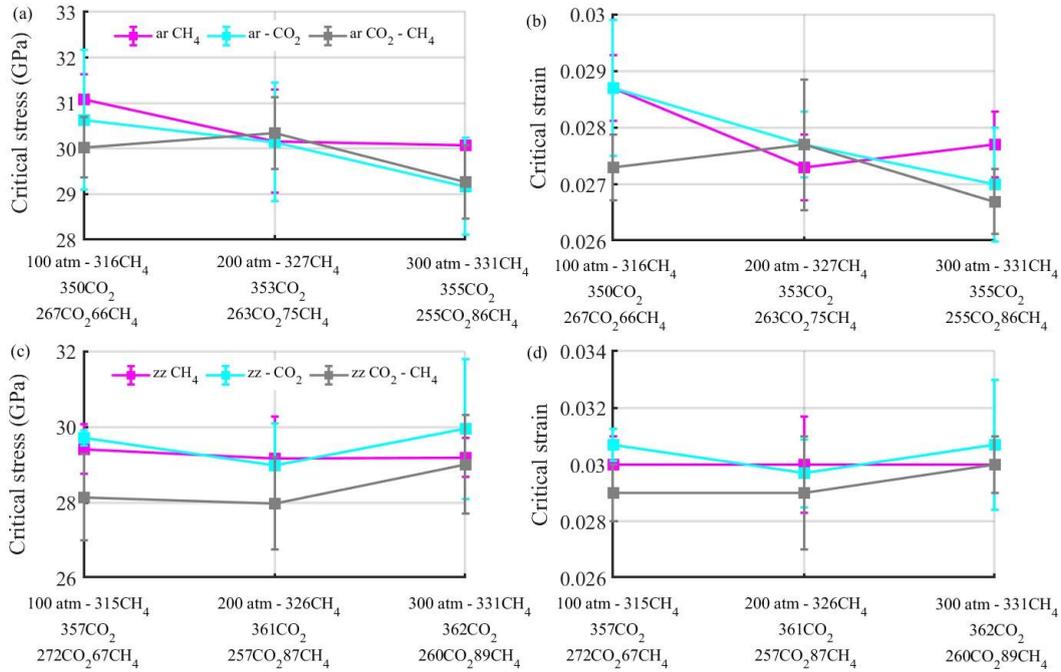

**Figure 5**: Comparison of critical stress and strain curves of (a) and (b) filled *ar* SWCNTs and (c) and (d) *zz* SWCNTs as a function of pressure and compositions. For 100 atm, the *ar* nanotube is filled with 316 molecules of $CH_4$, 350 $CO_2$, and 267 $CO_2$, and 66 $CH_4$ (mixture), while the *zz* nanotube is filled with 315 molecules of $CH_4$, 357 $CO_2$, and 272 $CO_2$, and 67 $CH_4$. The number of gas molecules for the other pressures are displayed on the x-axis. The magenta, cyan, and gray curves are for the $CH_4$, $CO_2$, and $CO_2/CH_4$ mixture, respectively.

We also investigated the influence of fillers on the nanotubes' failure. **Figure 6a** shows an empty *ar* nanotube under tension loading. The fracture starts on the bonds aligned to the loading direction, indicating that the covalent bonds play a central role in the nanotube's failure mechanism. The same failure is observed for the nanotubes filled with $CO_2$, as shown in **Fig. 6b.** We reach the same conclusion for the *zz* models (see **Fig. S5**). So, buckling modes are not affected by the presence of fillers in both directions (see **Figs. S6**), even for nanotubes of varying lengths or diameters (see **Fig. S7**). In addition, **Fig. S8** in the Supplementary material presents the $CO_2$ radial density profile within the nanotube during three different stages of compression loading, confirming that the arrangement of the $CO_2$ molecules has a minor effect on the buckling pattern.

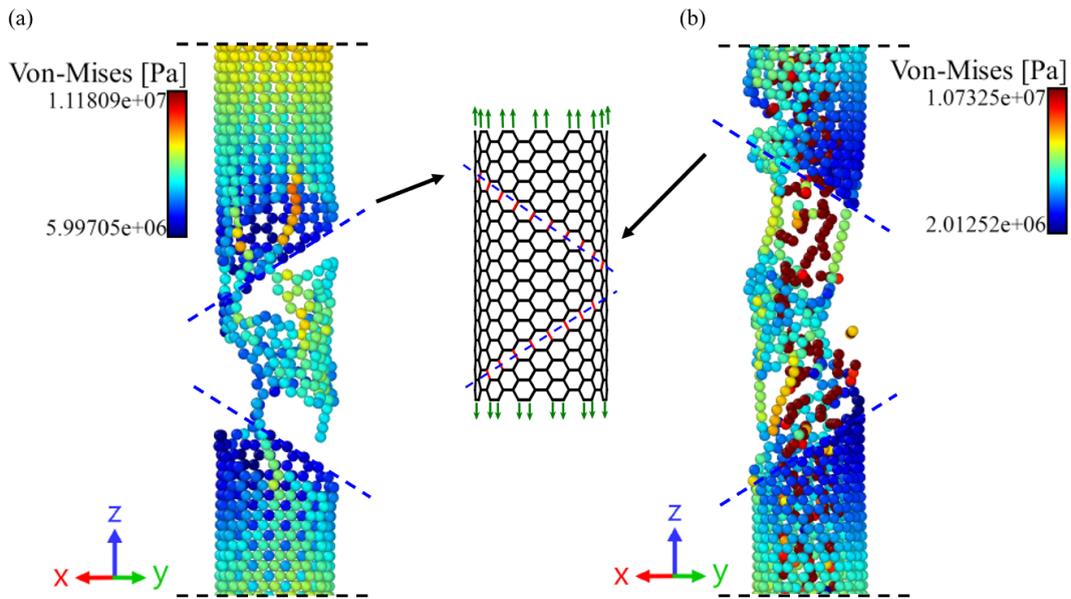

**Figure 6**: Fracture pattern for (a) empty and (b) filled *ar* SWCNTs under tension loading. The colors represent the potential energy distribution for each atom during the stretching process.

Moreover, the next results show that the mechanical performance of nanotubes is influenced by the relation between their length and diameter, regardless of the gas composition. To illustrate this effect, **Fig. 7** displays the critical strain for various aspect ratios (L/D = 5, L/D = 10, and L/D = 15), while **Fig. S9** shows the critical stress for the same models. For the nanotube with L/D = 5, a local buckling mode was observed, where the nanotube's axis remained straight with half waves along its wall, which resembles a thin-shell buckling, as defined by continuum mechanics approaches [63,64]. Higher L/D relations have sufficiently long lengths to display global buckling mode, similar to the beam models also described by theory at the macroscale [63]. We observed the local to global buckling modes transition occurred at an aspect ratio of 10. This transition from local to global modes was also observed in previous studies for empty (10,10) nanotubes [63]. These results highlight similarities in the buckling behavior of filled and unfilled nanotubes at both the nanometric and macroscopic scales.

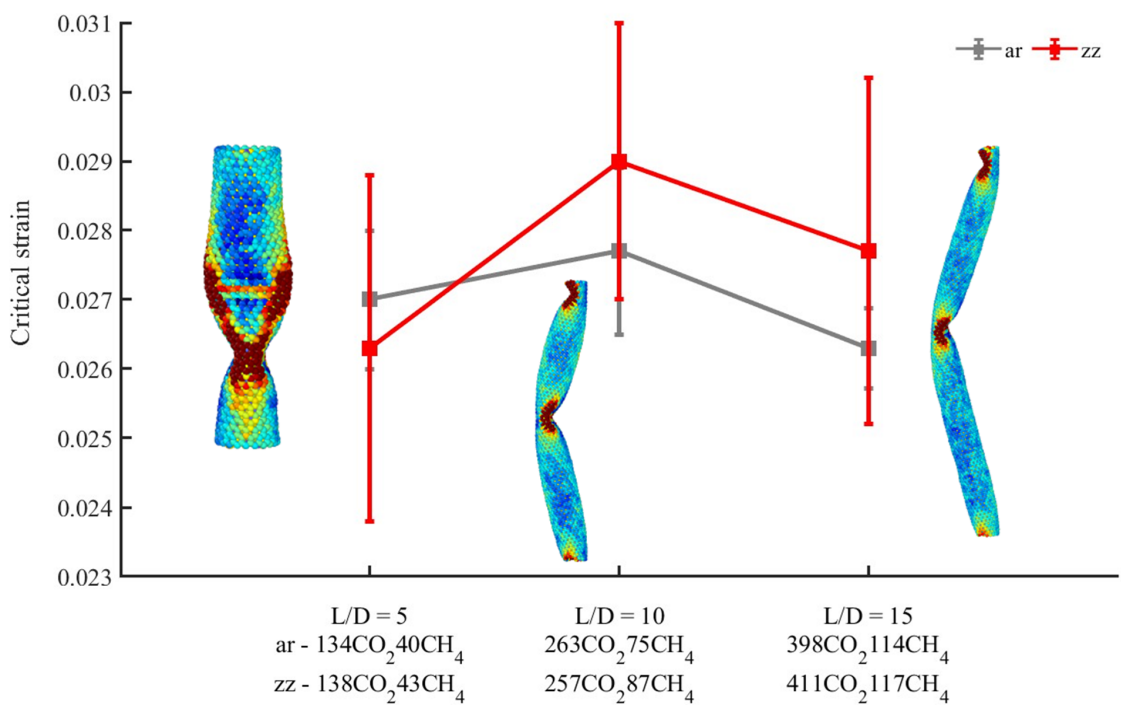

**Figure 7**: Critical strain and buckling modes for different aspect ratios and compositions. The colors represent the potential energy distribution for each atom during the stretching process. For L/D = 5, the *ar* nanotube is filled with 134 $CO_2$ molecules and 40 of $CH_4$, while the *zz* is filled with 138 $CO_2$ and 43 of $CH_4$. For L/D = 10, the *ar* nanotube is filled with 263 $CO_2$ molecules and 75 of $CH_4$, while the *zz* is filled with 257 $CO_2$ and 87 of $CH_4$. For L/D = 15, the *ar* SWCNT is filled with 398 $CO_2$ molecules and 114 of $CH_4$, while the *zz* is filled with 411 $CO_2$ and 117 $CH_4$.

## 4    Conclusions

We conducted a systematic study on the mechanical and adsorption properties of SWCNTs filled with greenhouse gases through GCMC and MD simulations using a recently developed parameterization for the LJ cross-terms. We evaluated how chirality, aspect ratio, pressure, and gas compositions impact the nanotubes' overall mechanical response under tensile and compressive strain.

A strong preference for the adsorption of $CO_2$ in comparison with $CH_4$ was observed, while for the smallest SWCNTs, namely, the (6,6) and (0,10), we observed that this preference is less evident. The fillers do not affect the mechanical response of nanotubes under tension and do not alter their failure pattern. However, gases significantly influence the mechanical performance of nanotubes under compression, with filled *zz* nanotubes exhibiting superior stability compared to *ar* ones. The study also identified similarities between the buckling of filled and unfilled nanotubes at different length scales.

The trends observed in our work indicate that the mechanical response of filled SWCNTs under compression has a complex relationship with the loading direction, nanotube diameters, chirality, and, to a minor extent, the gas composition. These findings confirm that all these features should be considered when designing technologies for gas capture and separation membranes; otherwise, the overall system may experience earlier failure in operational conditions. The study may guide the development of potential membrane-based devices for greenhouse gas separation under operational conditions.

**Acknowledgments**

The authors gratefully acknowledge the support of the RCGI – Research Centre for Gas Innovation, hosted by the University of São Paulo (USP) and sponsored by FAPESP – São Paulo Research Foundation (2014/50279-4, 2020/15230-5, and project number 2020/01558-9) and Shell Brasil, and the strategic importance of the support given by ANP (Brazil's National Oil, Natural Gas, and Biofuels Agency) through the R&D levy regulation. The authors also acknowledge the National Council for Scientific and Technological Development (CNPq) through grant 307064/2019-0 for financial support. The computational time for the calculations was provided by High-Performance Computing facilities at the University of de São Paulo (USP). HMC also thanks to the financial support of FAPESP, grants #2019/21430-0 and #2017/02317-2.

**Supporting Information: Mechanical and adsorption properties of greenhouse gases filled carbon nanotubes**

Daniela A. Damasceno[1], Henrique M. Cezar[2], Teresa D. Lanna[3], Alexsandro Kirch[4], Caetano R. Miranda[5*]

[1,3,4,5] Universidade de São Paulo, Instituto de Fisica, Rua do Matão 1371, São Paulo, SP 05508-090, Brazil

[2] Hylleraas Centre for Quantum Molecular Sciences and Department of Chemistry, University of Oslo, PO Box 1033 Blindern, 0315 Oslo, Norway

Corresponding author: *Caetano R. Miranda, email: crmiranda@usp.br

## S1 Interatomic potentials

To describe the intermolecular interactions, we used the LJ potential with a new set of parameters optimized for the $CO_2$-nanotube and $CH_4$-nanotube interfaces. The parameters are presented in **Table S1**. The LJ parameters were obtained using the Lorentz-Berthelot rules for the interaction between different molecule species.

**Table S1:** Crossing terms of the LJ interactions between the SWCNTs and the gases.

|  | $\epsilon_{ij}$ (K) | $\sigma_{ij}$ (Å) |
|---|---|---|
| $C_{SWCNT}$-$C_{CO2}$ | 74.466 | 3.09 |
| $C_{SWCNT}$-$O_{CO2}$ | 85.189 | 3.09 |
| $C_{SWCNT}$-$C_{CH4}$ | 66.540 | 3.38 |
| $C_{SWCNT}$-$H_{CH4}$ | 49.609 | 2.57 |

## S2 Grand Canonical Monte Carlo

**Table S2**: Ratio of $CO_2$:$CH_4$ molecules inside each studied SWCNT for the 20%$CO_2$:80%$CH_4$, 50%$CO_2$:50%$CH_4$, and 80%$CO_2$:20%$CH_4$ reservoirs at 200 atm, obtained from the GCMC simulations in **part I**. Values in parenthesis are the uncertainties propagated from the standard deviations of each species' average number of molecules in the GCMC simulations.

|  | Armchair | | | | | Zigzag | | | | |
|---|---|---|---|---|---|---|---|---|---|---|
| **Reservoir** | (6,6) | (8,8) | (10,10) | (12,12) | (14,14) | (0,10) | (0,14) | (0,17) | (0,21) | (0,24) |
| Mix 1 20%$CO_2$:80%$CH_4$ | 50:50 (1) | 85:15 (1) | 81:19 (1) | 78:22 (1) | 72:28 (1) | 49:51 (1) | 83:17 (3) | 78:22 (1) | 75:25 (1) | 74:26 (2) |
| Mix 2 50%$CO_2$:50%$CH_4$ | 55:45 (1) | 96:4 (1) | 90:10 (1) | 93:7 (1) | 89:11 (1) | 55:45 (1) | 92:8 (1) | 95:5 (2) | 92:8 (1) | 89:11 (1) |
| Mix 3 80%$CO_2$:20%$CH_4$ | 65:35 (4) | 97:3 (2) | 96:4 (2) | 97:3 (1) | 96:4 (1) | 60:40 (1) | 98:2 (1) | 97:3 (1) | 97:3 (1) | 95:5 (1) |

## S3 Molecular dynamics: Thermostat effects

It is known that thermostats in MD simulations may play an important role in the properties of confined systems. To investigate their effects, we considered two scenarios: in the first one, the thermostat was applied only on the nanotube walls, and in the second, on the nanotube and gases. In both cases, no significant difference was observed in the fracture strength and strain of the filled CNTs. Therefore, the thermostat was only applied to the nanotube walls in all simulations.

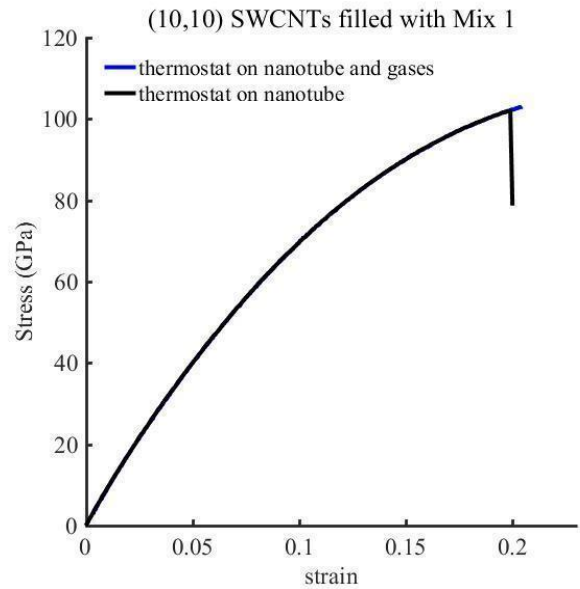

**Figure S1**: Stress-strain curves of (10,10) SWCNTs filled with Mix 1 and thermostat applied only on the nanotube walls (black curve) and on the nanotube and gases (blue curve). In both cases, no significant difference was observed in the fracture strength and strain of the filled CNTs. For the case where the thermostat was applied only on the nanotube walls, the fracture stress and strain were 102.1 GPa and 0.199, respectively, while for the other case, the fracture stress was 102.9 GPa and strain 0.204.

## S4 Results and discussions

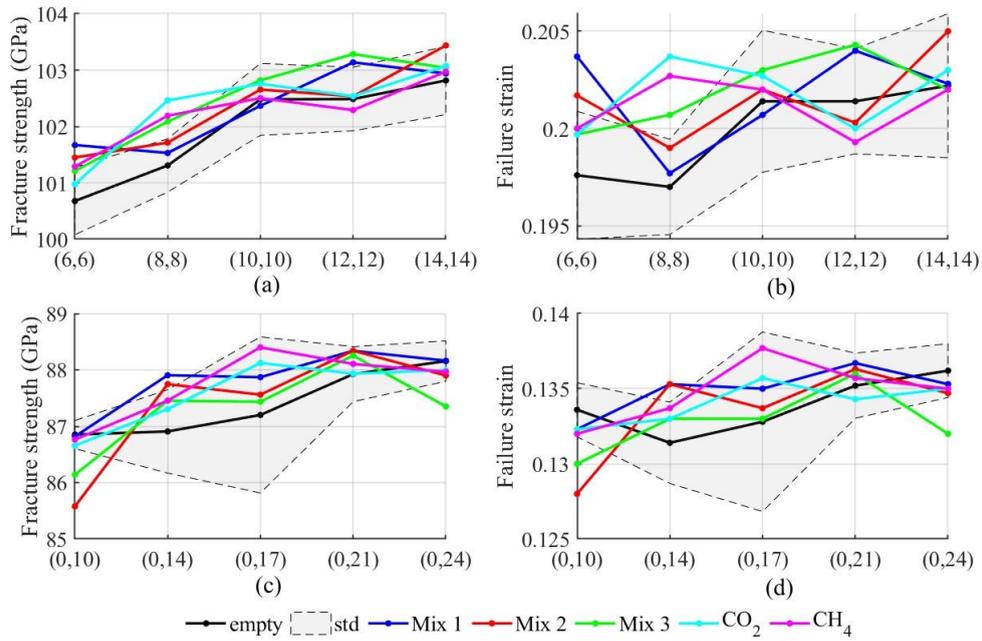

**Figure S2**: Comparison of (a) fracture strength and (b) failure strain of the empty and filled *ar* ((6,6), (8,8), (10,10), (12,12), and (14,14)); and (c) fracture strength and (d) failure strain of the empty and filled *zz* ((0,11), (0,14), (0,17), (0,21), and (0,24)) nanotubes. The mechanical response of empty nanotubes is represented by black curves and the gray region shows their standard deviation obtained based on an average of five different initial conditions. The blue, red, and green curves are for the compositions shown in **Table S2** (See Supporting Information section S2), while the cyan and magenta curves represent the pure compositions of $CO_2$ and $CH_4$, respectively.

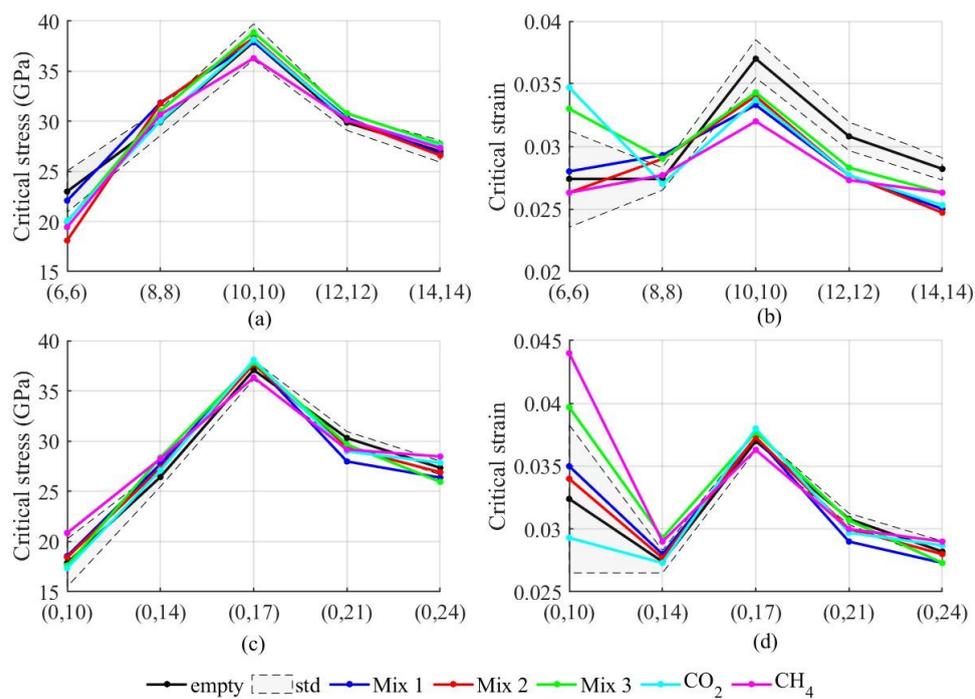

**Figure S3**: Comparison of critical (a) stress and (b) strain of the empty and filled *ar* ((6,6), (8,8), (10,10), (12,12), and (14,14)); and critical (c) stress and (d) strain of the empty and filled *zz* ((0,11), (0,14), (0,17), (0,21), and (0,24)) nanotubes. The mechanical response of empty nanotubes is represented by black curves and the gray region shows their standard deviation obtained based on an average of five different initial conditions. The blue, red, and green curves are for the compositions shown in **Table S2** (See Supporting Information section S2), while the cyan and magenta curves represent the pure compositions of $CO_2$ and $CH_4$, respectively.

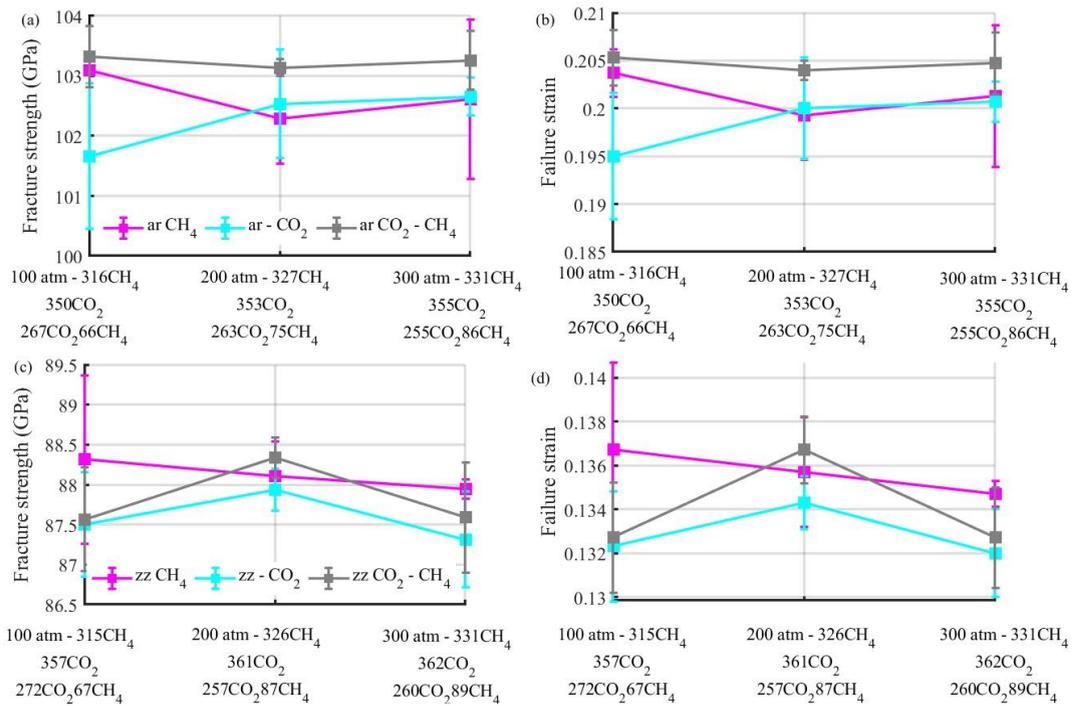

**Figure S4**: Comparison of Fracture strength and failure strain curves of (a) and (b) filled *ar* SWCNTs and (c) and (d) *zz* SWCNTs as a function of pressure and compositions. The number of gas molecules for each pressure is displayed on the x-axis. The cyan, magenta, and gray curves are for the $CO_2$, $CH_4$, and $CO_2/CH_4$ mixture.

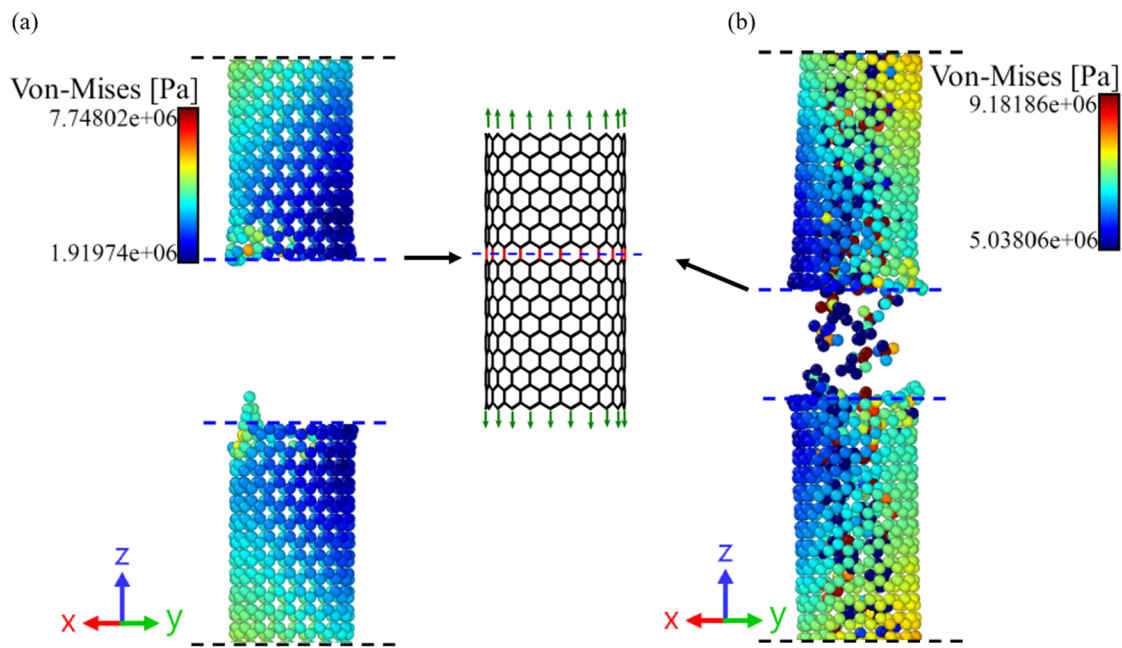

**Figure S5**: Fracture pattern for (a) empty and (b) filled zz SWCNTs.

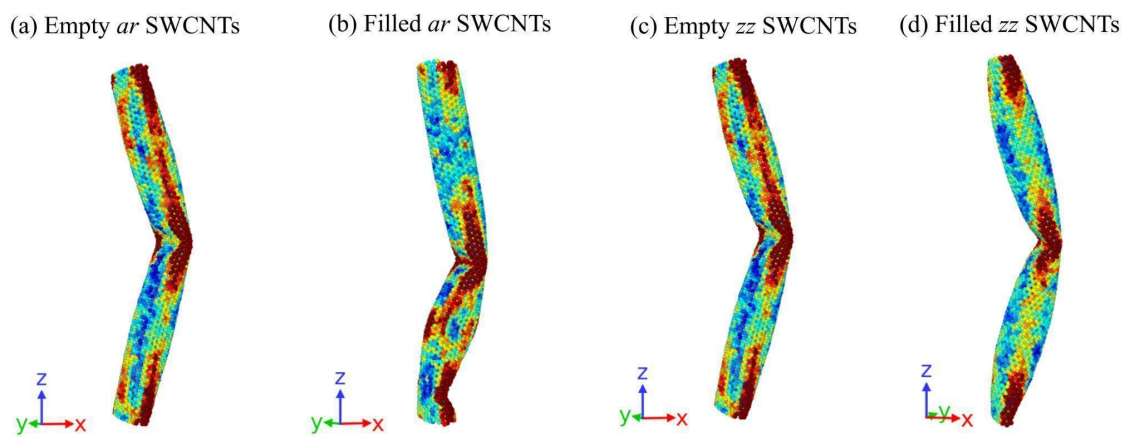

**Figure S6**: Buckling modes of (a) empty *ar* SWCNT, (b) filled *ar* SWCNT, (c) empty *zz* SWCNT, and (d) filled *zz* SWCNTs. The colors represent the potential energy distribution for each atom.

|  | under tension | under compression |
|---|---|---|
| ***ar* filled SWCNTs** | | |
| ***zz* filled SWCNTs** | | |

**Figure S7**: Fracture and buckling modes of smaller filled *ar* and *zz* SWCNTs.

**Figure S8** presents the $CO_2$ radial density profile within the (8,8) nanotube during three different stages of compression loading. The goal was to understand if the fillers may influence the buckling modes. The black curves correspond to the nanotube carbon atoms (represented by the gray atoms in the inset figure), and the blue and red curves correspond to the oxygen and carbon atoms of the $CO_2$, respectively. See the distribution of the $CO_2$ molecules in the inset figure. The number 1 in the caption represents the initial deformation stage from zero to 10%, followed by number 2 ranging from 10% to 20%, and number 3 from 20% to 30%. The x-axis corresponds to the nanotube's radii. The distribution of $CO_2$ molecules in the first and second stages is similar, while in the third case, the molecules are more spread. However, in the second stage, the nanotubes already started to buckle, showing that the arrangement of the $CO_2$ molecules did not change significantly to impact the buckling pattern.

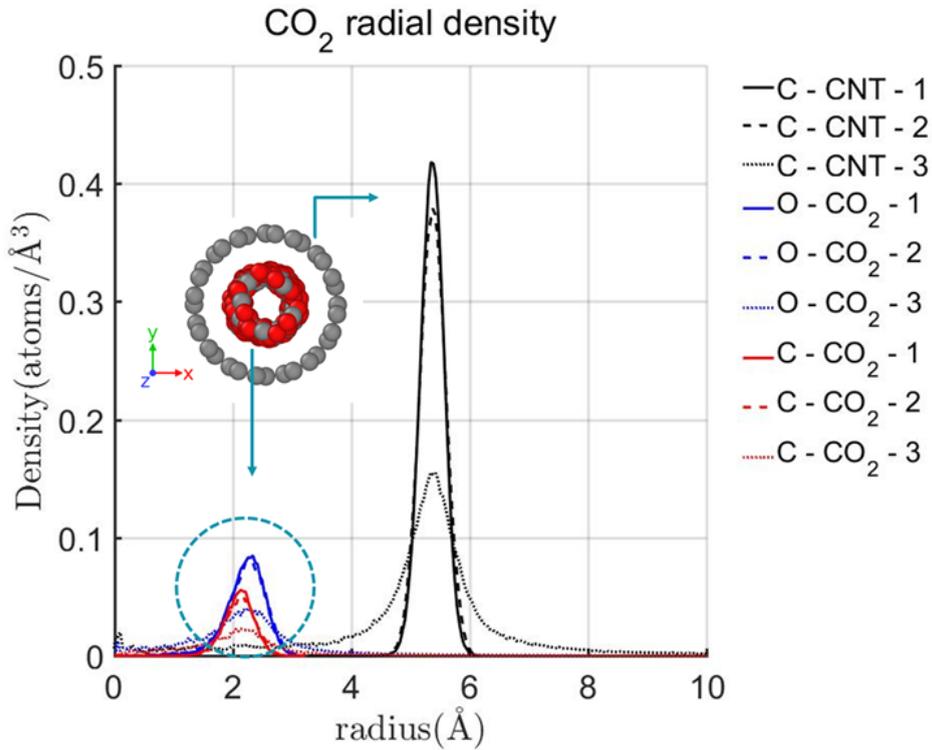

**Figure S8**: $CO_2$ radial density profile within the (8,8) nanotube. The black curves correspond to the nanotube's carbon, and the blue and red curves to the oxygen and carbon atoms of the $CO_2$, respectively. The numbers 1 – 3 represent the different stages of the deforming process.

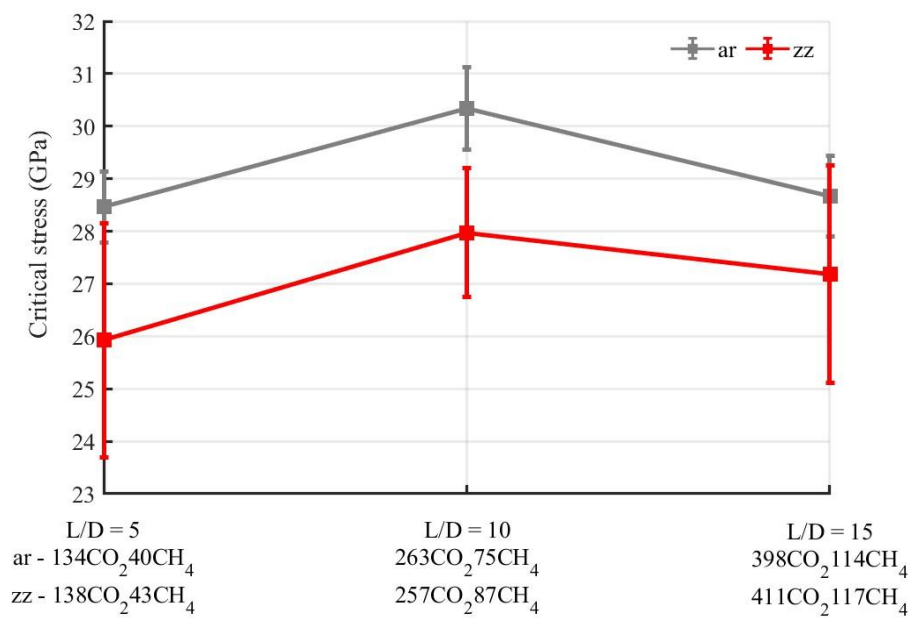

**Figure S9**: Critical stress of filled *ar* and *zz* SWCNTs for different aspect ratios. For L/D = 5, the *ar* nanotube is filled with 134 $CO_2$ molecules and 40 $CH_4$, while the *zz* is filled with 138 $CO_2$ and 43 $CH_4$. For L/D = 10, the *ar* nanotube is filled with 263 $CO_2$ molecules and 75 $CH_4$, while the *zz* is filled with 257 $CO_2$ and 87 $CH_4$. For L/D = 15, the *ar* SWCNT is filled with 398 $CO_2$ molecules and 114 $CH_4$, while the *zz* is filled with 411 $CO_2$ and 117 $CH_4$.